\documentclass[12pt]{article}
\usepackage{epsfig,amssymb,amsmath,graphicx,color}
\newcommand{\nn}{\nonumber}

\begin{document}

\vspace{9mm}

\begin{center}
{{{\Large \bf ${\cal N}=3$ Supersymmetric Effective Action of D2-branes\\
in Massive IIA String Theory}
}\\[17mm]
Gyungchoon Go${}^{1}$,~~O-Kab Kwon$^{1}$,~~D.~D. Tolla$^{1,2}$\\[3mm]
{\it $^{1}$Department of Physics,~BK21 Physics Research Division,
~Institute of Basic Science,\\
$^{2}$University College,\\
Sungkyunkwan University, Suwon 440-746, Korea}\\[2mm]
{\tt gcgo,~okab,~ddtolla@skku.edu} }
\end{center}

\vspace{20mm}

\begin{abstract}

We obtain a new-type of ${\cal N}=3$ Yang-Mills Chern-Simons theory from 
the Mukhi-Papageorgakis Higgsing of the ${\cal N}=3$ Gaiotto-Tomasiello theory.
This theory has ${\cal N}=1$ BPS fuzzy funnel solution which is expressed 
in terms of the seven generators of SU(3), excluding $T_8$.
We propose that this is an effective theory of multiple D2-branes with D6- and D8-branes 
background in massive IIA string theory.

\end{abstract}

\newpage

\tableofcontents

\section{Introduction}

Three dimensional Yang-Mills Chern-Simons (YM CS) theories  can be
realized on brane configurations in type II string theory in two
ways. On one hand, one can start with the Hanany-Witten type brane
configuration which contains D3-branes stretched between two
parallel NS5-branes in type IIB string theory~\cite{Hanany:1996ie}.
The corresponding gauge theory is  (2+1)-dimensional ${\cal N}=4$ YM
theory. When one of the NS5-brane is replaced by a (1,$k$)5-brane,
CS term with CS level $k$ arises in the corresponding gauge theories
and the supersymmetry is broken to ${\cal N}=1,2,3$ depending on the
orientation of the (1,$k$)5-brane with respect to the other
NS5-brane~\cite{Kitao:1998mf,Bergman:1999na}. For further progress on this issue, see \cite{Ohta:1999gj,Lee:1999ze,Kitao:1999uj,Ohta:1999iv}. This method
of generating the CS term was also used in the type IIB brane
configuration of the Aharony-Bergman-Jafferis-Maldacena (ABJM)
theory~\cite{Aharony:2008ug}, which describes the dynamics of
M2-branes on $\mathbb{C}^4/\mathbb{Z}_k$ orbifold singularity.
See also \cite{Hosomichi:2008jb,Aharony:2008gk,Aharony:2009fc}.
On the other hand, CS terms are also needed in describing D3(2)-branes
dynamics in the background of
D7(8)-branes~\cite{Green:1996bh,Bergshoeff:1996tu}. In this case the
CS term is generated by the monodromy due to the presence of the
D7-brane~\cite{Greene:1989ya} in type IIB brane configurations.
The corresponding CS term
in massive IIA brane configurations is obtained by {\it massive}
T-duality~\cite{Bergshoeff:1996ui}. This phenomenon is closely
related with the brane configuration of the Gaiotto-Tomasiello (GT)
theories~\cite{Gaiotto:2009mv,Gaiotto:2009yz}. In particular, by
introducing D7-branes to the type IIB brane configuration of ABJM
theory, Bergman and Lifschytz constructed the brane configuration of
the ${\cal N}=0$ GT theory~\cite{Bergman:2010xd}. For ${\cal N}=3$,
see \cite{Fujita:2009kw}.\footnote{Some aspects of ${\cal N}=2,3$ GT theories were also discussed in \cite{Aharony:2010af,Suyama:2010hr,Kwon:2011nv}.}

The dimensional reduction of the ABJM theory with U($N)\times$U($N$)
gauge symmetry~\cite{Aharony:2008ug} via the Mukhi-Papageorgakis
(MP) Higgsing procedure \cite{Mukhi:2008ux} results in the
$(2+1)$-dimensional ${\cal N}=8$ supersymmetric YM theory with
U($N$) gauge symmetry~\cite{Pang:2008hw,Kim:2011qv}. The ${\cal
N}=3$ GT theory~\cite{Gaiotto:2009mv} was obtained from the ABJM
theory by shifting the CS levels of the two gauge groups, so that
$k_1+k_2\ne 0$. Apparently, the ${\cal N}=3$ GT theory is a minor
deformation of the ABJM theory, however, there are unanswered
questions about this theory. This is mainly because of the fact that
there is no clear argument about the related brane system. In order
to clarify this point, we apply the MP Higgsing procedure to the
${\cal N}=3$ GT theory and obtain $(2+1)$-dimensional ${\cal N}=3$
YM CS theory with U$(N)$ gauge symmetry and CS level $k_1+k_2=F_0$. This
${\cal N}=3$ YM CS theory is different from the one studied
in~\cite{Kao:1993gs, Kao:1995gf} because it contains four massless
scalar fields and their fermionic superpartners in addition to the
three massive scalar fields in the massive vector multiplet, which
are also present in the latter theory. It is also true that our
theory has ${\cal N}=1$ BPS fuzzy funnel solution and $\left({\rm
R}^{4}\times S^1\right)^N/S_N$ vacuum moduli space while these are trivial in the theory in \cite{Kao:1993gs, Kao:1995gf}.

Even though the brane configuration for the original ${\cal N}=3$
GT theory is unclear, the structure of the moduli space and the
fuzzy funnel solution provide an insight in to the branes
configuration for our ${\cal N}=3$ YM CS theory. In this paper we
argue that in massive IIA string theory,
YM CS theories with U$(N)$
gauge symmetry describe the low energy dynamics of $N$ coincident
D2-branes in the background of D8-brane.\footnote{See
\cite{Brodie:2000ns,Imamura:2000jj} for earlier considerations in
the case of single D2-brane.} We also show that the presence of three massive and four massless scalar fields, which are matter contents of our ${\cal N}=3$ YM CS theory, implies
the branes system should contain D6-branes. More precisely, the
brane system includes $N$ coincident D2-branes in the background of
$|F_0|$ D8-branes, which have one common spacial direction with the
D2-branes, and $|F_0|$ D6-branes, which have two common spacial
directions with the D2-branes. The four massless scalar fields
represent the position of the D2-branes inside the worldvolume of
the D6-branes while the three massive scalar fields represent the
position of the D2-branes along the directions transverse to the
D6-branes in the presence of the background D8-branes.

The remaining part of the paper is organized as follows.
In section 2, we apply the MP Higgsing procedure to the ${\cal N}=3$ GT theory
and obtain ${\cal N}=3$ YM CS theory.
In section 3, we find the vacuum moduli space and ${\cal N}=1$ BPS fuzzy funnel solution.
In section 4, we propose the brane configuration for our ${\cal N}=3$ YM CS theory.
In section 5, we discuss our results and propose the type IIB brane configuration of the ${\cal N}=3$ GT theory.

\section{${\cal N}=3$ YM CS Theory}

\subsection{${\cal N}=3$ GT Theory}

Based on the ${\cal N}=2$ superfield formulation of
\cite{Gaiotto:2009mv}, the component field expansions of the GT
theories were obtained in~\cite{Kwon:2011nv}. For clarity of
presentation we copy the Lagrangian of the ${\cal N}=3$ GT
theory,
\begin{align}\label{N=3GT}
{\cal L}_{{\cal N}=3}={\cal L}_{0}+{\cal L}_{\rm CS}+{\cal L}_{\rm
ferm}+{\cal L}_{\rm bos},
\end{align}
where
\begin{align}
{\cal L}_{0}&={\rm tr}\big[-D_\mu Z^\dagger_AD^\mu Z^A-D_\mu
W^{\dagger A}D^\mu W_A+i\xi^{\dagger}_A\gamma^\mu
D_\mu\xi^A+i\omega^{\dagger A}\gamma^\mu D_\mu\omega_A\big],
\nonumber \\
{\cal L}_{\rm CS}&=\frac{k_1}{4\pi}\epsilon^{\mu\nu\rho} {\rm
tr}\Big(A_\mu\partial_\nu A_\rho+\frac{2i}3A_\mu A_\nu
A_\rho\Big)+\frac{k_2}{4\pi}\epsilon^{\mu\nu\rho} {\rm tr} \Big(\hat
A_\mu\partial_\nu \hat A_\rho +\frac{2i}3\hat A_\mu \hat A_\nu \hat
A_\rho\Big),
\nn \\
{\cal L}_{\rm ferm}&=-\frac{2\pi i}{k_1}{\rm tr}\Big[
(\xi^A\xi^\dagger_A-\omega^{\dagger
A}\omega_A)(Z^BZ^\dagger_B-W^{\dagger B}W_B) +2 (Z^A\xi^\dagger_A
-\omega^{\dagger A}W_A)(\xi^BZ^\dagger_B-W^{\dagger B}\omega_B)\Big]
\nonumber \\
&~~~-\frac{2\pi i}{k_2}{\rm
tr}\Big[(\xi^\dagger_A\xi^A-\omega_A\omega^{\dagger
A})(Z^{\dagger}_BZ^B-W_BW^{\dagger B})
+2(Z^\dagger_A\xi^A-\omega_AW^{\dagger A})(\xi^{\dagger}_BZ^B
-W_B\omega^{\dagger B})\Big]
\nonumber \\
&\hskip 1.5cm -\frac{2\pi}{k_1}\,{\rm
tr}\big(Z^A\omega_AZ^B\omega_B +\xi^AW_A\xi^BW_B
+2Z^AW_A\xi^B\omega_B +2Z^A\omega_A\xi^BW_B
\nonumber \\
&\hskip 1.5cm - \omega^{\dagger A}Z_A^\dagger\omega^{\dagger
B}Z_B^\dagger - W^{\dagger A}\xi_A^\dagger W^{\dagger
B}\xi_B^\dagger-2\omega^{\dagger A}\xi_A^\dagger W^{\dagger
B}Z_B^\dagger - 2 W^{\dagger A}\xi_A^\dagger\omega^{\dagger
B}Z_B^\dagger\big)
\nonumber \\
&~~~-\frac{2\pi}{k_2}\,{\rm
tr}\big(\omega_AZ^A\omega_BZ^B+W_A\xi^AW_B\xi^B
+2\omega_AZ^AW_B\xi^B +2W_AZ^A\omega_B\xi^B
\nonumber \\
&\hskip 1.5cm - Z_A^\dagger\omega^{\dagger
A}Z_B^\dagger\omega^{\dagger B} -\xi_A^\dagger W^{\dagger
A}\xi_B^\dagger W^{\dagger B} -2\xi_A^\dagger W^{\dagger A}
Z_B^\dagger\omega^{\dagger B} -2\xi_A^\dagger\omega^{\dagger A}
Z_B^\dagger W^{\dagger B}  \big),
\nn \\
{\cal L}_{{\rm bos}}
&= -\frac{4\pi^2}{k_1^2}{\rm tr}\Big[\big(Z^AZ_A^\dagger +
W^{\dagger A}W_A\big)\big(Z^BZ_B^\dagger - W^{\dagger B}W_B\big)
(Z^CZ_C^\dagger - W^{\dagger C}W_C\big)\Big]
\nonumber \\
&~~~-\frac{8\pi^2}{k_1k_2}{\rm tr}\Big[\big(Z^AZ_A^\dagger -
W^{\dagger A}W_A\big)Z^B (Z_C^\dagger Z^C - W_CW^{\dagger
C}\big)Z_B^\dagger
\nonumber \\
&~~~~~~~~~~~~~~~+\big(Z^AZ_A^\dagger - W^{\dagger
A}W_A\big)W^{\dagger B} (Z_C^\dagger Z^C - W_CW^{\dagger
C}\big)W_B\Big]
\nonumber \\
&~~~-\frac{4\pi^2}{k_2^2}{\rm tr}\Big[\big(Z_A^\dagger Z^A +
W_AW^{\dagger A}\big)\big(Z_B^\dagger Z^B - W_BW^{\dagger B}\big)
(Z_C^\dagger Z^C - W_CW^{\dagger C}\big)\Big]
\nonumber\\
&~~~ -4{\rm
tr}\Big[\big(\frac{2\pi}{k_1}W_AZ^BW_B + \frac{2\pi}{k_2} W_B Z^B
W_A\big) \big(\frac{2\pi}{k_1}W^{\dagger C}Z_C^\dagger W^{\dagger A}
+ \frac{2\pi}{k_2} W^{\dagger A}Z_C^\dagger W^{\dagger C}\big)
\nonumber \\
&~~~~~~~~~~~+\big(\frac{2\pi}{k_1}Z^B W_BZ^A + \frac{2\pi}{k_2} Z^A
W_B Z^B\big)\big(\frac{2\pi}{k_1} Z_A^\dagger W^{\dagger
C}Z_C^\dagger + \frac{2\pi}{k_2} Z_C^\dagger W^{\dagger C}
Z_A^\dagger\big)\Big].
\end{align}
In ${\cal N}=2$ superfield formalism $Z^A~{\rm and}~W_A~(A=1,2)$ are the scalar components of chiral superfields ${\cal Z}^A$ and ${\cal W}_A$, respectively, whereas $\xi^A$ and $\omega_A$ are their fermionic superpartners. $A_\mu$ and $\hat A_\mu$ are the vector components of the vector superfields ${\cal V}_1$ and ${\cal V}_2$, respectively.
The ${\cal N}=3$ supersymmetry transformation rules for these
component fields are as follows~\cite{Kwon:2011nv}:
\begin{align}\label{N=3susy}
&\delta Z^A = i\bar\epsilon\xi^A-\eta\omega^{\dagger A},
\quad \delta W_A = i\bar\epsilon \omega_A+ \eta\xi^\dagger_A,
\nonumber \\
&\delta\xi^A = -D_\mu Z^A \gamma^\mu\epsilon -\sigma_1Z^A\epsilon +
Z^A\sigma_2\epsilon -\frac{4\pi i}{k_1}\bar\epsilon W^{\dagger
B}Z_B^\dagger W^{\dagger A} -\frac{4\pi i}{k_2}\bar\epsilon
W^{\dagger A} Z_B^\dagger W^{\dagger B}
\nonumber \\
&\hskip 1.1cm +iD_\mu W^{\dagger A} \gamma^\mu\eta+ i\eta\sigma_1W^{\dagger A}
- i\eta W^{\dagger A}\sigma_2+\frac{4\pi i}{k_1}\eta W^{\dagger
B}Z_B^\dagger Z^A +\frac{4\pi i}{k_2}\eta Z^AZ_B^\dagger W^{\dagger
B},
\nonumber \\
 &\delta\omega_A = - D_\mu W_A\gamma^\mu\epsilon
+W_A\sigma_1\epsilon -\sigma_2 W_A\epsilon -\frac{4\pi
i}{k_1}\bar\epsilon Z_A^\dagger W^{\dagger B} Z_B^\dagger -
\frac{4\pi i}{k_2}\bar\epsilon Z_B^\dagger W^{\dagger B} Z_A^\dagger
\nonumber \\
&\hskip 1.1cm-iD_\mu Z^\dagger_A\gamma^\mu\eta+i\eta Z^\dagger_A\sigma_1
-i\eta\sigma_2 Z^\dagger_A-\frac{4\pi i}{k_1}\eta W_AW^{\dagger
B}Z^\dagger_B -\frac{4\pi i}{k_2}\eta Z^\dagger_BW^{\dagger B}W_A,
\nonumber \\
&\delta A_\mu = \frac12\big(\bar\epsilon\gamma_\mu\bar\chi_1 +
\chi_1\gamma_\mu\epsilon\big)-\frac12\big(\eta\gamma_\mu\zeta_1 -i
\bar\zeta_1\gamma_\mu\eta\big),\nonumber\\
& \delta\hat A_\mu =\frac12\big(\bar\epsilon\gamma_\mu\bar\chi_2 +
\chi_2\gamma_\mu\epsilon\big)+\frac12 \big(\eta\gamma_\mu\zeta_2 -i
\bar\zeta_2\gamma_\mu\eta\big),
\end{align}
where  $\epsilon$ and $\bar\epsilon$ are complex two component
spinor and its complex conjugate, whereas $\eta$ is a complex spinor
satisfying $\bar\eta=-i\eta$. Here we also defined
\begin{align}\label{auxilflds}
&\sigma_1 \equiv \frac{2\pi}{k_1}\big(Z^BZ_B^\dagger - W^{\dagger
B}W_B\big), \qquad ~\sigma_2 \equiv -\frac{2\pi}{k_2}\big(Z_B^\dagger
Z^B - W_BW^{\dagger B}\big),
\nonumber \\
&\chi_1 \equiv
-\frac{4\pi}{k_1}\big(Z^A\xi_A^\dagger-\omega^{\dagger A}W_A\big),
\qquad \chi_2 \equiv \frac{4\pi}{k_2} \big(\xi_A^\dagger Z^A -
W_A\omega^{\dagger A}\big),\nonumber\\
&\zeta_1\equiv \frac{4\pi}{k_1}\big(\xi^AW_A +
Z^A\omega_A\big),\qquad ~~~~\zeta_2\equiv\frac{4\pi}{k_2}\big(W_A\xi^A +
\omega_AZ^A\big).
\end{align}

In the next subsection we apply the MP Higgsing procedure to the
Lagrangian \eqref{N=3GT} and the corresponding supersymmetry
transformation rules \eqref{N=3susy} and obtain the ${\cal N}=3$ YM CS theory.

\subsection{MP Higgsing of the ${\cal N}=3$ GT Theory}

An important step in the MP Higgsing procedure is to turn on vacuum
expectation value $v$ for a scalar fields along which the bosonic
potential is flat. The only flat directions for the bosonic potential in
the ${\cal N}=3$ GT theory are the tilted directions, $Z^1\pm W^{\dagger 2}$ and $Z^2\pm W^{\dagger 1}$. In order to turn on the vacuum expectation value for a specific field, it is
convenient to make field redefinitions which align the scalars
along the flat directions of the potential. The appropriate field redefinitions for bifundamental fields are
\begin{align}\label{Fred2}
Z^A&=\frac{X^A-Y^{\dagger A}}{\sqrt 2},\quad W^{\dagger
A}=\frac{\sigma^A_{~B}(X^B+Y^{\dagger B})}{\sqrt
2},
\nn \\
\quad\xi^A&=\frac{\chi^A-i\lambda^{\dagger A}}{\sqrt 2},\quad ~~
\omega^{\dagger A}=\frac{\sigma^A_{~B}(\lambda^{\dagger
B}-i\chi^B)}{\sqrt 2},
\end{align}
where $\sigma^A_{~B}$ is the Pauli matrix $\sigma_1$. The ${\cal
N}=3$ GT Lagrangian is rewritten in terms of the redefined fields in appendix \ref{N=3GT2}.

The MP Higgsing procedure of the ABJM theory involves a double scaling
limit of large vacuum expectation value and CS level $k$, keeping the ratio $v/k$
finite. This can be applicable to the GT theory by setting $k_1=k$
and $k_2=-k+F_0$ and taking the same scaling limit.\footnote{In massive type IIA gravity, which is the gravity dual of the GT theory, $F_0$ is identified as the Romans mass~\cite{Romans:1985tz}.}
The appearance of the Chern-Simons levels in the fermionic and bosonic potentials suggests the following expansions in powers of $1/k$ for finite $F_0$,
\begin{align}\label{k2}
\frac1{k_2}&=-\frac1k\Big(1+\frac{F_0}k+\cdots\Big),
\nn \\
\frac1{k^2_2}&=\frac1{k^2}\Big(1+\frac{2F_0}k+\frac{3F_0^2}{k^2}+\cdots\Big),\nn\\
\frac1{k_1k_2}&=-\frac1{k^2}\Big(1+\frac{F_0}k+\frac{F_0^2}{k^2}+\cdots\Big).
\end{align}

We proceed by turning on the vacuum expectation value for a scalar field, which breaks the gauge symmetry from U($N$)$\times$U($N$) to U($N$) as follows:
\begin{align}\label{vev}
&X^A=\tilde X^{A}+i\tilde X^{A+4},\qquad Y^{\dagger A}=\frac
v2T^0\delta^{A2}+\tilde X^{A+2}+i\tilde X^{A+6},\nonumber\\
&\chi^A=\psi_{A}+i\psi_{A+4},\qquad ~~\lambda^{\dagger
A}=\psi_{A+2}+i\psi_{A+6}.
\end{align}
Here the fields $\tilde X^i(i=1,\cdots,8)$ and
$\psi_r(r=1,\cdots,8)$ are Hermitian and transform in the adjoint
representation of the unbroken U($N$) gauge group. In the double
scaling limit of $v,k\to\infty$ with finite $v/k$, the covariant
derivatives for the bosonic and fermionic fields are written as
\begin{align}
D_\mu Y^{\dagger 2}&=\tilde D_\mu \tilde X^4+iv\big(A_\mu^-+\frac
1v\tilde D_\mu \tilde X^8\big) \rightarrow\tilde D_\mu \tilde X^4+ivA_\mu^- ,
\nonumber\\
D_\mu Y^{\dagger 1}&=\tilde D_\mu \tilde X^{3}+i\tilde D_\mu\tilde
X^{7},\quad D_\mu X^A=\tilde D_\mu \tilde X^{A}+i\tilde D_\mu\tilde
X^{A+4},\nonumber\\
D_\mu \xi^A&=\tilde D_\mu \psi^{A}+i\tilde D_\mu\psi^{A+4},\quad
D_\mu \omega^{\dagger A}=\tilde D_\mu \psi^{A+2}+i\tilde
D_\mu\psi^{A+6},\label{covder}
\end{align}
where $A_{\mu}^\pm=\frac12(A_{\mu}\pm \hat A_{\mu})$, $\tilde D_\mu \tilde X = \partial_\mu \tilde X + i [A_\mu^+,\,\tilde X]$, and we have made the
gauge choice $A_\mu^-\to A_\mu^--\frac 1v\tilde D_\mu \tilde X^8$ in the first line.
 In writing \eqref{covder} we have also used the fact
 that the auxiliary field $A_\mu^-$ is of the order
$\frac 1v$ and neglected terms of this order or higher.

Using \eqref{vev} and \eqref{covder}, from the kinetic and the Chern-Simons terms in the ${\cal N}=3$ GT Lagrangian we obtain
\begin{align}
{\cal L}_0+{\cal L}_{\rm CS}= {\rm tr} \Big[&-\tilde D_\mu \tilde
X^i \tilde D^\mu \tilde X^i-v^2A_\mu^-A^{-\mu}+\frac k{2\pi}\
\epsilon^{\mu\nu\rho}A^-_\mu\tilde F_{\nu\rho}+
i\psi_r\gamma^\mu \tilde D_\mu\psi_r\nonumber\\
& +\frac{F_0}{4\pi}\epsilon^{\mu\nu\rho} \Big( A_\mu^+\partial_\nu
A_\rho^++\frac{2i}3A_\mu^+ A_\nu^+ A_\rho^+\Big)\Big] +{\cal
O}\big(\frac 1v\big),
\end{align}
where ${\tilde F}_{\mu\nu}=\partial_\mu A^+_\nu-\partial_\nu A^+_\mu
+ i [A^+_\mu,\, A^+_\nu]$ and $i=1,\cdots 7$ since $\tilde X^8$ is
eliminated by the gauge choice. Solving the equation of motion for the
auxiliary gauge field $A^-_{\mu}$ we can expresses it in terms of
the field strength of the dynamical gauge field $A^+_{\mu}$ as
\begin{align}
A^{-}_\mu =\frac k{4\pi v^2}\epsilon_\mu\!\!~^{\nu\rho} {\tilde
F}_{\nu\rho}=\frac{1}{2g v}\,\epsilon_\mu^{~\nu\rho} \tilde
F_{\nu\rho}, \label{Amu-}
\end{align}
where $g=\frac {2\pi v}k$ is the Yang-Mills coupling. For
dimensional reason it is also necessary to rescale all the matter
fields as $\phi\to\frac1g\phi$. Then we obtain
\begin{align}\label{L-YM-CS}
&{\cal L}_0+{\cal L}_{\rm CS}=\tilde{\cal L}_{\rm
YM} +\tilde{\cal L}_0+\tilde{\cal L}_{\rm CS}\nn\\
 &=\frac1{g^2}{\rm tr}
\Big[-\frac1{2}\tilde F_{\mu\nu}\tilde F^{\mu\nu}-\tilde D_\mu
\tilde X^i \tilde D^\mu \tilde X^i+ i\psi_r\gamma^\mu \tilde
D_\mu\psi_r +\frac{F_0g^2}{4\pi}\epsilon^{\mu\nu\rho} \Big(
A_\mu^+\partial_\nu A_\rho^++\frac{2i}3A_\mu^+ A_\nu^+
A_\rho^+\Big)\Big].
\end{align}

Using \eqref{k2} and \eqref{vev} the Higgsing of the potential terms
is tedious but straightforward. In particular, from the fermionic
potential we obtain Yukawa type coupling and fermionic mass
term, which are given by
\begin{align}
\tilde{\cal L}_{\rm ferm}={\rm tr}\Big(\frac
{iF_0}{4\pi}\mu^{rs}\psi_r\psi_s-\frac1{g^2}\Gamma_i^{rs}\psi_r[\tilde
X^i,\psi_s]\Big),
\end{align}
where $\Gamma_i^{rs}$'s are seven dimensional
Euclidian gamma matrices in a particular representation which is determined by the Higgsing procedure (see appendix \ref{Gam}),
and $\mu^{rs}$ is fermionic mass matrix given by
\[ \mu =\frac12 \left( \begin{array}{cccccccc}
-1&0&0&0&1&0&0&0 \\
0&1&0&0&0&1&0&0 \\
0&0&1&0&0&0&-1&0\\
0&0&0&1&0&0&0&1\\
1&0&0&0&-1&0&0&0\\
0&1&0&0&0&1&0&0\\
0&0&-1&0&0&0&1&0\\
0&0&0&1&0&0&0&1\end{array} \right).\]
For convenience we write
the Lagrangian in terms of the fermionic fields which are
eigenstates of this mass matrix. The mass matrix can be diagonalized
by an orthogonal matrix as follows
\begin{align}\label{ferm-mass}
\tilde\mu=O^{T}\mu O={\rm diag}(-1,1,1,1,0,0,0,0),
\end{align}
where $O$ is given by
\[ O =\frac1{\sqrt{2}} \left( \begin{array}{cccccccc}
-1&0&0&0&0&0&0&1 \\
0&0&0&1&0&0&-1&0 \\
0&0&-1&0&0&1&0&0\\
0&1&0&0&-1&0&0&0\\
1&0&0&0&0&0&0&1\\
0&0&0&1&0&0&1&0\\
0&0&1&0&0&1&0&0\\
0&1&0&0&1&0&0&0\end{array} \right).\]
Then the fermionic  mass eigenstates are
\begin{align}\label{tildepsi}
\tilde\psi_r&=\psi_sO^{sr}
\nn \\
&=\frac1{\sqrt{2}}(\psi_5-\psi_1,\,\psi_8
+\psi_4,\,\psi_7-\psi_3,\,\psi_6+\psi_2,\,\psi_8-\psi_4,\,\psi_7+\psi_3
,\,\psi_6-\psi_2,\,\psi_5+\psi_1).
\end{align}
This transformation also modifies the gamma matrices as
\begin{align}
\tilde\Gamma_i=O^{T}\Gamma_iO.
\end{align}
Then we can write
\begin{align}
\tilde{\cal L}_{\rm ferm}=\frac1{g^2}{\rm tr}\Big(\frac
{iF_0g^2}{4\pi}\tilde\mu^{rs}\tilde\psi_r\tilde\psi_s-\tilde\Gamma_i^{rs}\tilde\psi_r[\tilde
X^i,\tilde\psi_s]\Big).
\end{align}
The fermionic kinetic term in \eqref{L-YM-CS} is invariant under the
transformation \eqref{tildepsi}.

The Higgsing of the bosonic potential is even more involved  than
that of the fermionic potential, however, the algebraic procedure is
similar. As a result of such lengthy algebra we obtain\footnote{For
later convenience we have made renaming of scalar fields as follow:
$\tilde X^{1}\to\tilde X^{6}$, $\tilde X^{2}\to\tilde X^{3}$,
$\tilde X^{3}\to\tilde X^{4}$, $\tilde X^{4}\to\tilde X^{7}$,
$\tilde X^{5}\to\tilde X^{5}$, $\tilde X^{6}\to\tilde X^{8}$,
$\tilde X^{7}\to\tilde X^{9}$. The same renaming applies to the gamma matrices $\tilde\Gamma_i$.}
\begin{align}\label{bos-pot}
\tilde{\cal L}_{\rm bos}=\frac1{g^2}{\rm tr}\Big(-\frac
{F_0^2g^4}{16\pi^2}M_{ij}\tilde X^i\tilde
X^j-\frac{iF_0g^2}{2\pi}\tilde T_{ijk}\tilde X^i[\tilde X^j,\tilde
X^k]+\frac1{2}[\tilde X^i,\tilde X^j]^2\Big),
\end{align}
where the nonvanishing components of the bosonic mass matrix $M_{ij}$ and the antisymmetric constant tensor $\tilde T_{ijk}$ are
\begin{align}\label{bos-mass}
& M_{33}=M_{44}=M_{55} =1,\nonumber\\
&\tilde T_{567}=-\tilde T_{468}=\tilde T_{369}=\tilde T_{345}=\tilde
T_{378}=\tilde T_{479}=\tilde T_{589}=\frac16.
\end{align}

In summary, the total Lagrangian of the Higgsed theory is
written as
\begin{align}\label{SYMCS}
\tilde{\cal L}=\tilde{\cal L}_{{\rm YM}}+\tilde{\cal L}_{F_0},
\end{align}
where
\begin{align}
\tilde{\cal L}_{{\rm YM}}&=\frac1{g^2}{\rm tr}
\Big(-\frac1{2}\tilde F_{\mu\nu}\tilde F^{\mu\nu}-\tilde D_\mu
\tilde X^i \tilde D^\mu \tilde X^i+ i\psi_r\gamma^\mu \tilde
D_\mu\psi_r-\tilde\Gamma_i^{rs}\tilde\psi_r[\tilde
X^i,\tilde\psi_s]+\frac1{2}[\tilde X^i,\tilde X^j]^2\Big),
\\
\tilde{\cal L}_{F_0}&=\frac{F_0}{4\pi}{\rm tr} \Big(
\epsilon^{\mu\nu\rho} \big( A_\mu^+\partial_\nu
A_\rho^++\frac{2i}3A_\mu^+ A_\nu^+
A_\rho^+\big)+i\tilde\mu^{rs}\tilde\psi_r\tilde\psi_s -2i\tilde T_{ijk}\tilde
X^i[\tilde X^j,\tilde X^k] -\frac
{F_0g^2}{4\pi}M_{ij}\tilde X^i\tilde X^j\Big).\nn
\end{align}
This is the ${\cal N}=3$ YM CS theory anticipated at the end of the
previous subsection. For vanishing $F_0$ this reduces to ${\cal
N}=8$ super YM theory as expected. In literature (2+1)-dimensional
${\cal N}=3$ YM CS  theory was already studied \cite{Kao:1993gs,Kao:1995gf}.
In this case the theory can be obtained from the  ${\cal N}=4$ YM
theory by adding CS term, which breaks one supersymmetry. The field
contents of the later differ from the field contents our ${\cal
N}=3$ YM CS theory by four massless scalars and their superpartners.
The Lagrangian of \cite{Kao:1993gs,Kao:1995gf} can also be obtained
by turning off four scalar fields $\tilde X^{6,7,8,9}$ and four
Majorana fermions $\tilde\psi_{5,6,7,8}$ in our YM CS Lagrangian.

The supersymmetry transformation rules of \eqref{SYMCS} are obtained
as a result of the Higgsing of the corresponding transformation
rules in the original GT theory given in \eqref{N=3susy}
\begin{align}\label{SUSY-rules}
&\delta A^+_{\mu}=i\epsilon_{r}\gamma_{\mu}\tilde\psi_r,\qquad
\delta \tilde X^i=i\tilde\Gamma_i^{rs}\epsilon_r
\tilde\psi_s,\nonumber\\
&\delta\tilde\psi_r= i\tilde
F_{\mu\nu}\sigma^{\mu\nu}\epsilon_r+\tilde\Gamma_i^{rs}\gamma^{\mu}\epsilon_s\tilde
D_\mu\tilde X^i -\tilde\Gamma_{ij}^{rs}\epsilon_{s}[\tilde
X^i,\tilde
X^j]-\frac{F_0g^2}{4\pi}\tilde\mu^{rs}\tilde\Gamma_i^{st}\epsilon_t\tilde
X^i,
\end{align}
where the nonvanishing supersymmetry parameters are
\begin{align}\label{susy-para}
\epsilon_2=-\frac{1+i}{2\sqrt 2}(\bar\epsilon-
i\epsilon),\quad\epsilon_3=-\frac{1-i}{\sqrt
2}\eta,\quad\epsilon_4=\frac{1-i}{2\sqrt 2}(\bar\epsilon+
i\epsilon),
\end{align} and
\begin{align}
\sigma^{\mu\nu}=-\frac i4\big(\gamma^\mu\gamma^\nu
-\gamma^\nu\gamma^\mu\big),\quad\quad \tilde\Gamma_{ij}=\frac
i4\big(\tilde\Gamma_i\tilde\Gamma_j-\tilde\Gamma_j\tilde\Gamma_i\big).
\end{align}
Actually, the Higgsing of \eqref{N=3susy} gives the supersymmetric transformation
rules of the dynamical fields, which are the seven scalar fields
$\tilde X^i$, the eight fermionic fields $\tilde\psi_r$, and the
gauge field $A_\mu^+$, as well as the transformation rules for the
auxiliary gauge field $A_\mu^-$, and the would-be Goldstone boson
$\tilde X^8$. However, the fields $A_\mu^-$ and ${\tilde X}^8$ are
integrated out from the action and their transformation rules, which
are not listed in \eqref{SUSY-rules}, are irrelevant.

\section{Vacuum Moduli Space and Fuzzy Funnel Solution}\label{vmsffs}

\subsection{Vacuum moduli space}

In order to understand the brane configuration for our ${\cal N}=3$
YM CS theory in \eqref{SYMCS}, we start by figuring out the vacuum
moduli space of the theory.
The bosonic potential in \eqref{bos-pot} can be written in a positive-definite form
as follows:
\begin{align}
V_{{\rm bos}} = \frac1{4g^2}\sum_{r=1}^{8}\left|\left((1-i)\tilde\Gamma_{ij}^{r2} -(1+i)\tilde\Gamma_{ij}^{r4}\right)[\tilde X^i,\,\tilde X^j] + \beta\tilde\mu^{rs}
\left((1-i)\tilde\Gamma_{i}^{s2} -(1+i)\tilde\Gamma_{i}^{s4}\right)\tilde X^i\right|^2,
\end{align}
where $\displaystyle{\beta\equiv\frac{F_0g^2}{4\pi}}$ and we have introduced the notation
$|{\cal O}|^2 \equiv {\rm tr} {\cal O}^\dagger {\cal O}$.
We obtain the vacuum equations from this positive-definite potential,
\begin{align}\label{veq}
&[\tilde X^{a},\, \tilde X^{b}]=0,\quad [\tilde X^{a},\, \tilde X^{p}]=0,
\nn \\
&\beta \tilde X^3 + i([\tilde X^6,\tilde X^9]+[\tilde X^7,\tilde X^8])=0,
\nn \\
&\beta \tilde X^4 - i([\tilde X^6,\tilde X^8]-[\tilde X^7,\tilde X^9])=0,
\nonumber\\
&\beta \tilde X^5 + i([\tilde X^6,\tilde X^7]+[\tilde X^8,\tilde X^9])=0,
\end{align}
where  $a,b=3,4,5$, $p=6,7,8,9$.
The solution of \eqref{veq} is
\begin{align}
\tilde X^a =0, \qquad \tilde X^p = diagonal\,\,matrices.
\end{align}
The diagonal matrices $\tilde X^p$'s represent the full moduli space
of the theory. The fact that the $N\times N$ scalar fields are
diagonal on the vacuum moduli indicates that the U($N$) gauge
symmetry of the theory is broken to ${\rm U}(1)^N\times S_N$, where
the $S_N$ permutes the diagonal elements of the matrices. Thus the
moduli space including the effect of the dual photon in (2+1)-dimensions is given by \begin{align}\label{vacmod}
{\cal M} = \frac{\left({\rm R}^{4}\times S^1\right)^N}{S_N}.
\end{align}

\subsection{Fuzzy funnel solution}
In this subsection, we will obtain fuzzy funnel solution of BPS equations in our $\cal N$=3 YM CS theory.
The Killing spinor equation of the supersymmetry variation \eqref{SUSY-rules} is written as
\begin{equation}\label{Keq}
\delta\tilde\psi_r= i\tilde
F_{\mu\nu}\sigma^{\mu\nu}\epsilon_r+\tilde\Gamma_i^{rs}\gamma^{\mu}\epsilon_s\tilde
D_\mu\tilde X^i -\tilde\Gamma_{ij}^{rs}\epsilon_{s}[
\tilde X^i,
\tilde X^j]-\beta\tilde\mu^{rs}\tilde\Gamma_i^{st}\epsilon_t
\tilde X^i=0.
\end{equation}

In order to obtain a fuzzy funnel solution, we consider the following projection to the supersymmetry parameters, $\gamma^2 \epsilon_{2,3}=\epsilon_{2,3}$ and also set $\epsilon_{4}=0$. The resulting fuzzy funnel solution
reduces the number of supersymmetries by $1/3$, i.e., it has ${\cal N}=1$ supersymmetry. We also assume the vanishing gauge field and a static configuration.
Under these conditions, the BPS equations are
\begin{align}\label{BPSeq}
&\tilde\Gamma_i^{rs}\partial_1 \tilde X^i=0,
\nn\\
&\tilde\Gamma_i^{rs} \partial_2
\tilde X^i-\tilde\Gamma_{ij}^{rs} [\tilde X^i,\tilde X^j]-\beta \mu^{rt}\tilde\Gamma_i^{ts} \tilde X^i=0.
\end{align}

The first line of \eqref{BPSeq} can be satisfied by choosing a configuration which does not depend on $x_1$ direction.
From the second line of \eqref{BPSeq} we have
\begin{align}\label{BPSeq2}
&\partial_2 \tilde X^3-i[\tilde X^4,\tilde X^5]=0,\qquad \partial_2 \tilde X^5-i[\tilde X^3,\tilde X^4]=0,\nonumber\\
&\partial_2 \tilde X^4-\beta \tilde X^4
+i([\tilde X^6,\tilde X^8]+[\tilde X^3,\tilde X^5]-[\tilde X^7,\tilde X^9])=0,\nonumber\\
&\beta
\tilde X^3+i([\tilde X^6,\tilde X^9]+[\tilde X^7,\tilde X^8])=0,\qquad \beta \tilde X^5+i([\tilde X^6,\tilde X^7]+[\tilde X^8,\tilde X^9])=0,\nonumber\\
&\partial_2 \tilde X^6-i[\tilde X^4,\tilde X^8]=0,\qquad \partial_2
\tilde X^7+i[\tilde X^4,\tilde X^9]=0,\nonumber\\
&\partial_2 \tilde X^8-i[\tilde X^6,\tilde X^4]=0,\qquad
\partial_2
\tilde X^9-i[\tilde X^4,\tilde X^7]=0,\nonumber\\
&[\tilde X^6,\tilde X^5]-[\tilde X^3,\tilde X^8]=0,\qquad [\tilde X^6,\tilde X^3]+[\tilde X^5,\tilde X^8]=0,\nonumber\\
&[\tilde X^3,\tilde X^7]+[\tilde X^5,\tilde X^9]=0,\qquad [\tilde X^3,\tilde X^9]+[\tilde X^7,\tilde X^5]=0.
\end{align}
Comparing the equation in the second line with the remaining equations, it appears natural to divide it into the following two equations
\begin{equation}
\beta \tilde X^4-i([\tilde X^6,\tilde X^8]-[\tilde X^7,\tilde X^9])=0,\qquad \partial_2 \tilde X^4+i[\tilde X^3,\tilde X^5]=0.
\end{equation}
Then from the first and the second lines of \eqref{BPSeq2} we obtain
\begin{align}\label{fuzzy}
\partial_2 \tilde X^a=i\epsilon^{abc}[\tilde X^b,\tilde X^c],\qquad (a,b,c=3,4,5).
\end{align}
These are the Nahm equations with the fuzzy two sphere solution, in which the scalar fields $\tilde X^{3,4,5}$ are proportional to the generators of SU(2).
However, the fuzzy two sphere configuration does not satisfy the remaining equations in \eqref{BPSeq2}.
It is also important to notice that there is no nontrivial solution satisfying
the equations \eqref{BPSeq2} in the case of U(2) gauge group.
For $N\ge3$, an interesting solution exists and it can be expressed in terms of
seven generators of SU(3). Explicitly, we can write
\begin{align}\label{eqg}
\tilde X^3=g(x_2)\, T_1,\qquad \tilde X^4=g(x_2)\, T_2, \qquad \tilde X^5=g(x_2)\, T_3,
\end{align}
where $T_{1,2,3}$'s are the SU(2) subgroup elements of
$N$-dimensional representation of SU(3). Then from \eqref{fuzzy} we
easily obtain
\begin{equation}\nn
g(x_2)=\frac{1}{x_2}.
\end{equation}
The remaining equations of \eqref{BPSeq2} can be solved by choosing $\tilde X^{6,7,8,9}$ in terms of the rest of generators of SU(3), excluding $T_8$,
\begin{align}\label{eqh}
\tilde X^6=h(x_2)\,T_4,\qquad \tilde X^7=h(x_2)\,T_5,\qquad \tilde X^8=-h(x_2)\,T_6,\qquad
\tilde X^9=h(x_2)\,T_7,
\end{align}
where
\begin{align}\nn
h(x_2)=\pm\sqrt{\beta \, g(x_2)}=\pm\sqrt{\frac{\beta}{x_2}} \,.
\end{align}
Here we would like to point out an important difference between our
${\cal N}=3$ YM CS theory and a similar theory in
\cite{Kao:1993gs,Kao:1995gf}. As we pointed out before, the
latter theory can be obtained from ours by turning off four massless
scalar fields, $\tilde X^{6,7,8,9}$, which means in that case the
fuzzy funnel solution in \eqref{eqg} and \eqref{eqh} is not allowed
for nonvanishing $\beta$. As we will see in the next section,
together with the vacuum moduli space, this ${\cal N}=1$ BPS
solution provides useful insights about the brane configuration of
our theory.

\section{Brane Configuration}\label{braneconf}

\subsection{Generation of CS terms}
In order to pave a way for the understanding of the brane
configuration, which can be described by our YM CS theory obtained in section 2,
we briefly summarize some brane configurations in the literature.
These brane configurations are described by gauge theories involving CS terms.
We start with a type IIB brane system where two parallel
NS5-branes separated along one direction of the worldvolume of
$N$ D3-branes.
The remaining two worldvolume coordinates of the
D3-branes are parallel to the corresponding coordinates of
NS5-branes. In the low energy limit, this configuration is described by
(2+1)-dimensional ${\cal N}=4$ YM theory with gauge group U($N$)~\cite{Hanany:1996ie}, where all fields transform in the adjoint
representations. Since the two NS5-branes are parallel, the three
scalar fields, representing the positions of the D3-branes inside
the worldvolume of NS5-branes, are massless.

Now we replace one of the NS5-branes with a (1,$k$)5-brane (a bound
state of an NS5-brane and $k$ D5-branes) in a tilted direction with respect to the other NS5-brane. Then the D3-branes cannot move
freely and this fact translates into mass terms for the three scalar
fields on the field theory side. The ${\cal N}=4$ supersymmetry of
the original theory is broken to ${\cal N}=1,2,3$ theories,
depending on the choice of the direction of the (1,$k$)5-brane. The
corresponding effective field theories for these cases are obtained
by including the CS terms with CS level $k$ in supersymmetric
ways~\cite{Kitao:1998mf,Bergman:1999na}. Such CS term is introduced
in order to cancel the surface term originated from the boundary
condition of the (1,$k$)5-brane in the equation of motion of the
gauge field~\cite{Kitao:1998mf,Bergman:1999na}.

The brane configuration of the ABJM theory~\cite{Aharony:2008ug} is
based on the brane system of the ${\cal N}=3$ YM CS
theory~\cite{Kao:1993gs,Kao:1995gf}. An important difference is the
fact that the two parallel NS5-branes are separated along a compact
direction of the worldvolume of $N$ D3-branes. In this case, the
D3-branes, which wind around the compact direction, can break on the
NS5-branes resulting in a (2+1)-dimensional ${\cal N}=3$ YM CS gauge
theory with gauge group ${\rm U}_{k}(N)\times{\rm
U}_{-k}(N)$~\cite{Aharony:2008ug}. At the infrared fixed point, this
becomes conformal and the supersymmetry is enhanced to ${\cal N}=6$.
One can also add $l$ fractional D3-branes, suspended on one side of
the interval between the NS5-brane and the (1,$k$)5-brane. Then the
corresponding effective field theory becomes ${\cal N}=3$ YM CS
theory with gauge group ${\rm U}(N+l)_k\times{\rm U}(N)_{-k}$ or
${\rm U}(N)_k\times{\rm U}(N+l)_{-k}$ depending on the side on which
the fractional D3-branes are added~\cite{Aharony:2008gk}.

Along a different line of thought, CS terms are also required in order
to describe brane systems involving D7- or
D8-branes~\cite{Green:1996bh,Bergshoeff:1996tu,Bergman:2010xd}.
The configuration with
D8-branes can be understood by the {\it massive} T-dualization of that of
D7-branes~\cite{Bergshoeff:1996ui}. In \cite{Bergman:2010xd}, a
D7-brane was added to the brane configuration of the ABJM theory as
follows:
\begin{center}
\begin{tabular}{c|cccccccccc}
& 0 & 1 & 2 & 3 & 4 & 5 & 6 & 7 & 8 & 9 \\
\hline
$N$ D3 & $\bullet$ & $\bullet$ & $\bullet$ & & & & $\bullet$ & & &\\
1 D7 & $\bullet$ & $\bullet$ & $\bullet$ & $\bullet$&$\bullet$ & &  &$\bullet$ &$\bullet$ &$\bullet$
\end{tabular}
\end{center}
Here we have omitted 5-branes for simplicity. This configuration
breaks the entire supersymmetry. Since the D7-brane is a pointlike
object in the ($x_5,x_6$)-plane, it sources a SL(2,$\mathbb{Z}$)
monodromy on the plane, $\tau\to\tau+1$,\footnote{We define the complex
combination of the axion field $C_0$ and the dilaton field $\phi$ as
$\tau \equiv \frac{C_0}{2\pi} + i e^{-\phi}$.} i.e., $C_0\to
C_0+2\pi$ for the axion. This monodromy is the result of a branch
cut emanated from the D7-brane with the direction of the cut chosen
to cross the D3-branes. Then the Wess-Zumino type coupling for the
D3-branes generates a CS term:
\begin{align}
&\int_{{\rm R}^{2+1}}\int_{x_6} C_0 {\rm tr}(F\wedge F)  \sim S_{{\rm R}^{2+1}}^{{\rm CS}}(A).
\end{align}

To summarize, we have seen two ways to generate the CS term in  the
descriptions of brane configurations in (2+1)-dimensional gauge
theories. The CS term in our ${\cal N}=3$ YM CS theory is related to
the configuration involving D7- or D8-branes. In the next
subsection, we construct the brane configuration for our ${\cal
N}=3$ YM CS theory starting with type IIB brane system involving
D7-branes.

\subsection{Massive IIA brane configuration}\label{IIAbr}

The type IIA string theory on $AdS_4\times \mathbb{CP}^3$ with $q$ D8-branes ($q=|F_0|$)
wrapped on $\mathbb{CP}^3$ was proposed as a dual gravity of the ${\cal N}=3$ GT theory~\cite{Fujita:2009kw}. Based on this and the type IIB brane configuration of the
${\cal N}=6$ ABJM theory, we propose the type IIB brane configuration
of the ${\cal N}=3$ GT theory as follows:
\begin{center}
\begin{tabular}{c|cccccccccc}
& 0 & 1 & 2 & 3 & 4 & 5 & $\hat 6$ & 7 & 8 & 9 \\
\hline
$N$ D3 & $\bullet$ & $\bullet$ & $\bullet$ & & & & $\bullet$ & & &\\
NS5 & $\bullet$ & $\bullet$ & $\bullet$ & $\bullet$ & $\bullet$ & $\bullet$ & & &  \\
$(1,k)5$ & $\bullet$ & $\bullet$ & $\bullet$ & $\cos\theta$ &
$\cos\theta$ &$\cos\theta$ & & $\sin\theta$ & $\sin\theta$ &
$\sin\theta$
\\
$q$ D5 & $\bullet$ & $\bullet$ & $\bullet$ &  & &  & &$\bullet$ &$\bullet$ &$\bullet$  \\
$q$ D7 & $\bullet$ & $\bullet$ &  & $\bullet$ & $\bullet$ &$\bullet$
& & $\bullet$ & $\bullet$ & $\bullet$
\end{tabular}
\end{center}
\begin{center}{
\small Table 1: The NS5-brane, $q$ D5-branes, and $q$ D7-branes are located at the same point along the $x_6$-direction.}
\end{center}
where $\hat 6$ represents a compact direction and $\theta$ is the orientation of the (1,$k$)5-brane
relative to NS5-brane in ($x_3$,$x_7$)-, ($x_4$,$x_8$)-, and
($x_5$,$x_9$)-planes, and  $\tan\theta=k$, assuming the string
coupling $g_s=1$ and RR axion is vanishing.
In addition to the brane configuration of the ABJM theory,
this configuration contains $q$ D7-branes and additional $q$ D5-branes
in a supersymmetric way. The D7-branes are results of the T-dualization of the D8-branes
in the proposal of \cite{Fujita:2009kw}, while
the additional D5-branes are included in our proposed brane configuration for the reason
that we will explain below.

The MP Higgsing procedure in ABJM theory includes two
important steps, which are identification of the two gauge fields with each
other and moving the M2-branes far away from the orbifold
singularity. In the corresponding brane configuration these actions
are interpreted as  separating the D3-branes from the five-branes
and moving them far away in the transverse directions. After the
separation, the T-duality along the compact direction will give the
brane configuration with coincident D2-branes, and the corresponding
effective field theory is the ${\cal N}=8$ YM theory in (2+1)-dimensions.
This procedure does not break supersymmetry.

Even though the M-theory
uplifting of our proposed brane configuration is unclear, we can apply the
MP Higgsing procedure to this brane configuration as well.
This corresponds to moving the NS5- and (1,$k$)5-brane far away from
the D7-D3-D5-brane system in the transverse directions.
This results in the type IIB brane configuration with $N$ D3-branes intersecting $q$ D7-branes along one common spatial direction and $q$ D5-branes along two common spatial directions as follows:
\begin{center}
\begin{tabular}{c|cccccccccc}
& 0 & 1 & 2 & 3 & 4 & 5 & $\hat 6$ & 7 & 8 & 9 \\
\hline
$N$ D3 & $\bullet$ & $\bullet$ & $\bullet$ & & & & $\bullet$ & & &\\
$q$ D5 & $\bullet$ & $\bullet$ & $\bullet$ &  & &  & &$\bullet$ &$\bullet$ &$\bullet$  \\
$q$ D7 & $\bullet$ & $\bullet$ &  & $\bullet$ & $\bullet$ &$\bullet$ &
& $\bullet$ & $\bullet$ & $\bullet$
\end{tabular}
\end{center}
\begin{center}{
\small Table 2: Type IIB brane configuration for ${\cal N}=3$ GT theory after the MP Higgsing.}
\end{center}
Unlike the brane configuration in \cite{Bergman:2010xd} ours is supersymmetric.
Based on the discussion in the previous subsection, $q$ D7-branes generate CS term with CS level $\pm q$ depending on the relative orientation of D3- and D7-branes.

Applying {\it massive} T-duality along $x_{\hat 6}$-direction from IIB configuration with D7-branes to massive IIA configuration with D8-branes~\cite{Bergshoeff:1996ui}, we obtain the following brane configuration:
\begin{center}
\begin{tabular}{c|cccccccccc}
& 0 & 1 & 2 & 3 & 4 & 5 & 6 & 7 & 8 & 9 \\
\hline
$N$ D2 & $\bullet$ & $\bullet$ & $\bullet$ & & & &  & & &\\
$q$ D6 & $\bullet$ & $\bullet$ & $\bullet$ &  & &  & $\bullet$&$\bullet$ &$\bullet$ &$\bullet$  \\
$q$ D8 & $\bullet$ & $\bullet$ &  & $\bullet$ & $\bullet$ &$\bullet$ &$\bullet$
& $\bullet$ & $\bullet$ & $\bullet$
\end{tabular}
\end{center}
\begin{center}{
\small Table 3: Massive IIA brane configuration for ${\cal N}=3$ YM CS theory}
\end{center}
where $6$ denotes the new direction appeared after the T-dualization
along $x_{\hat 6}$-direction. This brane configuration
is expected to coincide with the brane configuration described
by the ${\cal N}=3$ CS YM theory discussed in section 2.

Next we use the vacuum moduli space and the fuzzy funnel solution in section 3 to discuss
the importance of D6-branes in the brane configuration of massive IIA string theory
in Tab.3. From the vacuum moduli space in \eqref{vacmod} we can infer that there are three massive directions for which $\langle\tilde X^{3,4,5}\rangle=0$ and four flat directions for which $\langle \tilde X^{6,7,8,9}\rangle=diagonal$. The former indicates the fact that the D2-branes are not free to move in these directions, while they are free to move in the remaining four transverse directions. This moduli space and the supersymmetry structure in the
massive IIA gravity~\cite{Fujita:2009kw} suggest the presence of D6-branes parallel to
the D2-branes in addition to D8-branes.
Moreover, the ${\cal N}=1$ BPS fuzzy funnel solution, in which the seven transverse scalar fields are proportional to the seven generators (excluding $T_8$) of SU(3) with $x_2$-dependent coefficients, also seems to support our brane configuration.
The solution is given by $\tilde X^{3,4,5}\sim (1/{x_2})T_{1,2,3}$ and $\tilde X^{6,7,8,9}\sim (1/\sqrt{x_2})T_{4,5,6,7}$. The $(1/x_2)$-dependence of $\tilde X^{3,4,5}$ indicates the localization of the D8-branes along those directions without any interference from the D6-branes. On the other hand the $(1/\sqrt{x_2})$-dependence of $\tilde X^{6,7,8,9}$ indicates mild localization of the D8-branes along those directions due to an interference from the D6-branes which span the $x_2$-direction.
Further evidence for this brane configuration should come from the BPS solutions
in the massive IIA gravity. We leave this possibility for future investigation.

\section{Conclusion}

In this paper we carried out the MP Higgsing of the ${\cal N}=3$ GT
theory and obtained ${\cal N}=3$ YM CS theory in (2+1)-dimensions
with U$(N)$ gauge symmetry. We also verified that the MP Higgsing of
the supersymmetry transformation rules of the GT theory results in
the corresponding rules in the YM CS theory. Compared to the MP
Higgsing of the ABJM theory, the present case is more subtle because
of two reasons. First, non of the four complex scalars in the GT
theory represent the flat direction of the bosonic potential and
they can not acquire infinitely large vacuum expectation values.
We overcame this problem by introducing field redefinitions which
rotate the scalars to the flat directions of the bosonic potential.
Second, in the GT theory we have two CS levels $k_1$ and $k_2$ and
it is not clear how to take the large CS level limit. We
took $k_1,k_2\to \pm\infty$ limit under the assumption that $k_1+k_2=F_0$ and $F_0$ is a finite dimensionless parameter.
It turns out that the $F_0$ is the CS
level in the resulting YM CS theory.

Earlier, ${\cal N}=3$ YM CS theory was studied from different
perspective~\cite{Kao:1993gs, Kao:1995gf}. This theory is a
deformation of the ${\cal N}=4$ YM theory in (2+1)-dimensions by a
CS term. On the other hand, our ${\cal N}=3$ YM CS theory is a
similar deformation of the ${\cal N}=8$ YM theory in
(2+1)-dimensions. By comparing these two theories, one can realize
that the former  is obtained from the latter by turning off four
massless scalars and their fermionic superpartners. An interesting
difference between these two theories is the fact that in our theory
we could find a non trivial fuzzy funnel solutions to the BPS
equations while in their theory such BPS solution does not exist. In
addition, the vacuum moduli space in our theory is $\left({\rm
R}^{4}\times S^1\right)^N/S_N$,  while it is trivial in their theory.

Since the ${\cal N}=3$ YM CS theory we obtained in this paper is
new, we found it interesting to figure out the brane configuration
which can be described by this theory. We proposed that the theory
describes the dynamics of $N$ coincident D2-branes in the background
of $q$ D6-branes and the same number of D8-branes, $q$ being the
absolute value of the CS level $F_0$. More precisely, the branes
system contains $N$ D2-branes extending along the directions
$x_{0,1,2}$, $q$ D6-branes along the directions $x_{0,1,2,6,7,8,9}$,
and $q$ D8-branes along the directions $x_{0,1,3,4,5,6,7,8,9}$.
As a confirmation of our brane configuration, we obtained
${\cal N}=1$ BPS fuzzy funnel solution which indicates the
localization of the D8-branes along the $x_2$-direction and supports the presence
of D6-branes.

The massive IIA supergravity~\cite{Romans:1985tz}
is the low energy limit of the massive IIA string theory. This supergravity theory has many (non)supersymmetric solutions of the form ${\rm AdS}_4\times {\cal M}_6$~\cite{Romans:1985tz,Behrndt:2004mj,Lust:2004ig,Tomasiello:2007eq,Koerber:2008rx,
Petrini:2009ur,Lust:2009mb,Aharony:2010af,Tomasiello:2010zz}, where ${\cal M}_6$ represents a six dimensional manifold.
The supersymmetries of these solutions are less than ${\cal N}=3$.
Since our massive IIA brane configuration in subsection \ref{IIAbr} has
${\cal N}=3$ supersymmetry, finding the corresponding solution in gravity side
will be interesting.

\section*{Acknowledgements}
The authors would like to thank Min-Young Choi, Shinsuke Kawai, Yoonbai Kim,  Sangmin Lee, Corneliu Sochichiu, Takao Suyama, and Tadashi Takayanagi for
helpful discussions. This work was supported by the Korea Research
Foundation Grant funded by the Korean Government  with grant number
2011-0009972 (O.K.) and 2009-0077423 (D.D.T.).

\appendix

\section{${\cal N}=3$ GT Lagrangian after Field Redefinition}\label{N=3GT2}

After the field redefinition \eqref{Fred2}, the ${\cal N}=3$ GT Lagrangian in \eqref{N=3GT} is rewritten  as
\begin{align}
{\cal L}_{0}&={\rm tr}\big[-D_\mu X^\dagger_AD^\mu X^A-D_\mu
Y^{\dagger A}D^\mu Y_A+i\chi^{\dagger}_A\gamma^\mu
D_\mu\chi^A+i\lambda^{\dagger A}\gamma^\mu D_\mu\lambda_A\big],
\nonumber \\
{\cal L}_{\rm CS}&=\frac{k_1}{4\pi}\epsilon^{\mu\nu\rho} {\rm
tr}\Big(A_\mu\partial_\nu A_\rho+\frac{2i}3A_\mu A_\nu
A_\rho\Big)+\frac{k_2}{4\pi}\epsilon^{\mu\nu\rho} {\rm tr} \Big(\hat
A_\mu\partial_\nu \hat A_\rho +\frac{2i}3\hat A_\mu \hat A_\nu \hat
A_\rho\Big),
\nn \\
{\cal L}_{\rm ferm}&=\frac{2\pi }{k_1}{\rm tr}\Big[(\lambda^{\dagger
A}\chi^\dagger_A-\chi^A\lambda_A)(X^BY_B+Y^{\dagger
B}X^\dagger_B)\nonumber\\
&~~~~~~~~~~~+\frac12\big(\lambda^{\dagger
A}X^\dagger_A-i\chi^AY_A-iX^A\lambda_A+Y^{\dagger
A}\chi^\dagger_A\big)\big(\lambda^{\dagger
B}X^\dagger_B-i\chi^BY_B-iX^B\lambda_B+Y^{\dagger
B}\chi^\dagger_B\big)\nonumber\\
&~~~~~~~~~~~+\frac12\big(\chi^{A}X^\dagger_A+i\lambda^{\dagger
A}Y_A-Y^{\dagger
A}\lambda_A-iX^A\chi^\dagger_A\big)\big(\chi^{B}X^\dagger_B+i\lambda^{\dagger
B}Y_B-Y^{\dagger
B}\lambda_B-iX^B\chi^\dagger_B\big)\nonumber\\
&~~~~+\sigma^A_{~C}\sigma^B_{~D}\big\{(\chi^C\lambda_A+\lambda^{\dagger
C}\chi^\dagger_A)(Y^{\dagger
D}X^\dagger_B-X^DY_B)+i(\lambda^{\dagger
C}\lambda_A-\chi^C\chi^\dagger_A)(X^{D}X^\dagger_B-Y^{\dagger D}Y_B)\nonumber\\
&~~~~~~~~~~-\frac12(\chi^CX^\dagger_A-i\lambda^{\dagger
C}Y_A-Y^{\dagger
C}\lambda_A+iX^C\chi^\dagger_A)(\chi^DX^\dagger_B-i\lambda^{\dagger
D}Y_B-Y^{\dagger D}\lambda_B+iX^D\chi^\dagger_B)\nonumber\\
&~~~~~~~~~~+\frac12(\lambda^{\dagger
C}X^\dagger_A+i\chi^CY_A+iX^C\lambda_A+Y^{\dagger
C}\chi^\dagger_A)(\lambda^{\dagger
D}X^\dagger_B+i\chi^DY_B+iX^D\lambda_B+Y^{\dagger
D}\chi^\dagger_B)\big\}\Big]
\nonumber \\
&+\frac{2\pi }{k_2}{\rm tr}\Big[(\chi^\dagger_A\lambda^{\dagger
A}-\lambda_A\chi^A)(Y_BX^B+X^\dagger_BY^{\dagger
B})\nonumber\\
&~~~~~~~~+\frac12\big(X^\dagger_A\lambda^{\dagger
A}-iY_A\chi^A-i\lambda_AX^A+\chi^\dagger_AY^{\dagger
A}\big)\big(X^\dagger_B\lambda^{\dagger
B}-iY_B\chi^B-i\lambda_BX^B+\chi^\dagger_BY^{\dagger
B}\big)\nonumber\\
&~~~~~~~~+\frac12\big(X^\dagger_A\chi^{A}+iY_A\lambda^{\dagger
A}-\lambda_AY^{\dagger
A}-i\chi^\dagger_AX^A\big)\big(X^\dagger_B\chi^{B}+iY_B\lambda^{\dagger
B}-\lambda_BY^{\dagger
B}-i\chi^\dagger_BX^B\big)\nonumber
\end{align}
\begin{align}
&~~~~+\sigma_A^{~C}\sigma_B^{~D}\big\{(\lambda_C\chi^A+\chi^\dagger_C\lambda^{\dagger
A})(X^\dagger_DY^{\dagger B}-Y_DX^B)+i(\lambda_C\lambda^{\dagger
A}-\chi^\dagger_C\chi^A)(X^\dagger_DX^B-Y_DY^{\dagger B})\nonumber\\
&~~~~~~~~~-\frac12(X^\dagger_C\chi^A-iY_C\lambda^{\dagger
A}-\lambda_CY^{\dagger
A}+i\chi^\dagger_CX^A)(X^\dagger_D\chi^B-iY_D\lambda^{\dagger
B}-\lambda_DY^{\dagger B}+i\chi^\dagger_DX^B)\nonumber\\
&~~~~~~~~~+\frac12(X^\dagger_C\lambda^{\dagger
A}+iY_C\chi^A+i\lambda_CX^A+\chi^\dagger_CY^{\dagger
A})(X^\dagger_D\lambda^{\dagger
B}+iY_D\chi^B+i\lambda_DX^B+\chi^\dagger_DY^{\dagger B})\big\}\Big],\nn
\end{align}
and
\begin{align}\label{LbosDF}
{\cal L}_{{\rm bos}} &= -\frac{4\pi^2}{k_1^2}{\rm
tr}\Big[\big(X^AX_A^\dagger + Y^{\dagger A}Y_A\big)\big(X^BY_B +
Y^{\dagger B}X^\dagger_B\big) (X^CY_C+ Y^{\dagger
C}X^\dagger_C\big)\nonumber\\
&~~~~~~~~+\frac12\sigma_B^{~D}\sigma^C_{~E}(X^BX^\dagger_D+X^BY_D-Y^{\dagger
B}X^\dagger_D-Y^{\dagger B}Y_D)\nonumber\\
&~~~~~~~~~~~~\times(X^AX^\dagger_A-X^AY_A-Y^{\dagger
A}X^\dagger_A+Y^{\dagger A}Y_A)(X^EX^\dagger_C-X^EY_C+Y^{\dagger
E}X^\dagger_C-Y^{\dagger E}Y_C)\nonumber\\
&~~~~~~~~+\frac12\sigma_B^{~D}\sigma^C_{~E}(X^\dagger_AX^B+Y_AX^B-X^\dagger
_AY^{\dagger B}-Y_AY^{\dagger B})\nonumber\\
&~~~~~~~~~~~~\times(X^\dagger_DX^E+Y_DX^E+X^\dagger _DY^{\dagger
E}+Y_DY^{\dagger E})(X^\dagger_CX^A-Y_CX^A+X^\dagger _CY^{\dagger
A}-Y_CY^{\dagger A})\Big]
\nonumber \\
&~~~-\frac{4\pi^2}{k_2^2}{\rm tr}\Big[\big(X_A^\dagger X^A +
Y_AY^{\dagger A}\big)\big(X_B^\dagger Y^{\dagger B} +Y_BX^B\big)
(X_C^\dagger Y^{\dagger C} + Y_CX^C\big)\nonumber\\
&~~~~~~~~+\frac12\sigma_B^{~D}\sigma^C_{~E}(X^\dagger_DX^B+Y_DX^B-X^\dagger_DY^{\dagger
B}-Y_DY^{\dagger B})\nonumber\\
&~~~~~~~~~~~~\times(X^\dagger_AX^A+Y_AX^A+X^\dagger_AY^{\dagger
A}+Y_AY^{\dagger A})(X^\dagger_CX^E-Y_CX^E+X^\dagger_CY^{\dagger
E}-Y_CY^{\dagger E})\nonumber\\
&~~~~~~~~+\frac12\sigma_B^{~D}\sigma^C_{~E}(X^AX^\dagger_D+X^AY_D-Y^{\dagger
A}X^\dagger_D-Y^{\dagger A}Y_D)\nonumber\\
&~~~~~~~~~~~~~\times(X^BX^\dagger_C-X^BY_C-Y^{\dagger
B}X^\dagger_C+Y^{\dagger B}Y_C)(X^EX^\dagger_A-X^EY_A+Y^{\dagger
E}X^\dagger_A-Y^{\dagger E}Y_A)\Big]
\nonumber \\
&~~~-\frac{8\pi^2}{k_1k_2}{\rm tr}\Big[\big(X^AY_A + Y^{\dagger
A}X^\dagger_A\big)\big\{X^B (X_C^\dagger Y^{\dagger C} +
Y_CX^C\big)X_B^\dagger +Y^{\dagger B} (X_C^\dagger Y^{\dagger C} +
Y_CX^C\big)Y_B\big\}
\nonumber\\
&~~~~~~~~+\frac14\sigma_B^{~D}\sigma^C_{~E}(X^BX^\dagger_D+X^BY_D-Y^{\dagger
B}X^\dagger_D-Y^{\dagger B}Y_D)
\nonumber\\
&~~~~~~~~~~~~\times(X^AX^\dagger_C-X^AY_C-Y^{\dagger
A}X^\dagger_C+Y^{\dagger A}Y_C)(X^EX^\dagger_A-X^EY_A+Y^{\dagger
E}X^\dagger_A-Y^{\dagger E}Y_A)
\nonumber\\
&~~~~~~~~+\frac14\sigma_B^{~D}\sigma^C_{~E}(X^\dagger_AX^B+Y_AX^B-X^\dagger
_AY^{\dagger B}-Y_AY^{\dagger B})
\nonumber\\
&~~~~~~~~~~~~\times(X^\dagger_DX^A+Y_DX^A+X^\dagger _DY^{\dagger
A}+Y_DY^{\dagger A})(X^\dagger_CX^E-Y_CX^E+X^\dagger _CY^{\dagger
E}-Y_CY^{\dagger E})
\nonumber\\
&~~~~~~~~+\frac14\sigma_B^{~D}\sigma^C_{~E}(X^\dagger_DX^B+Y_DX^B-X^\dagger_DY^{\dagger
B}-Y_DY^{\dagger B})
\nonumber\\
&~~~~~~~~~~~~\times(X^\dagger_AX^E+Y_AX^E+X^\dagger_AY^{\dagger
E}+Y_AY^{\dagger E})(X^\dagger_CX^A-Y_CX^A+X^\dagger_CY^{\dagger
A}-Y_CY^{\dagger A})\nonumber\\
&~~~~~~~~+\frac14\sigma_B^{~D}\sigma^C_{~E}(X^AX^\dagger_D+X^AY_D-Y^{\dagger
A}X^\dagger_D-Y^{\dagger A}Y_D)
\nonumber\\
&~~~~~~~~~~~~~\times(X^BX^\dagger_A-X^BY_A-Y^{\dagger
B}X^\dagger_A+Y^{\dagger B}Y_A)(X^EX^\dagger_C-X^EY_C+Y^{\dagger
E}X^\dagger_C-Y^{\dagger E}Y_C)\Big].
\end{align}

\section{Seven Dimensional Euclidean Gamma Matrices}\label{Gam}

In subsection 2.2 we have seen that the MP Higgsing of the fermionic potential gives the fermionic mass term and Yukawa-type coupling which is expressed in terms of seven dimensional Euclidean Gamma matrices.
In this appendix we list the the Gamma matrices which were determined by the Higgsing procedure,

\footnotesize{
\begin{align}
\begin{array}{ll}
\Gamma_1 = \left(\begin{array}{cccccccc}
0&0&0&1&0&0&0&0\\
0&0&0&0&0&0&1&0\\
0&0&0&0&0&-1&0&0\\
-1&0&0&0&0&0&0&0\\
0&0&0&0&0&0&0&1\\
0&0&1&0&0&0&0&0\\
0&-1&0&0&0&0&0&0\\
0&0&0&0&-1&0&0&0
\end{array}\right),
& \Gamma_2=\left(\begin{array}{cccccccc}
0&0&-1&0&0&0&0&0\\
0&0&0&0&0&0&0&1\\
1&0&0&0&0&0&0&0\\
0&0&0&0&0&-1&0&0\\
0&0&0&0&0&0&-1&0\\
0&0&0&1&0&0&0&0\\
0&0&0&0&1&0&0&0\\
0&-1&0&0&0&0&0&0
\end{array}\right),\\
\Gamma_3 = \left(\begin{array}{cccccccc}
0&0&0&0&0&1&0&0\\
0&0&0&0&1&0&0&0\\
0&0&0&1&0&0&0&0\\
0&0&-1&0&0&0&0&0\\
0&-1&0&0&0&0&0&0\\
-1&0&0&0&0&0&0&0\\
0&0&0&0&0&0&0&-1\\
0&0&0&0&0&0&1&0
\end{array}\right),
& \Gamma_4=\left(\begin{array}{cccccccc}
0&0&0&0&-1&0&0&0\\
0&0&0&0&0&1&0&0\\
0&0&0&0&0&0&1&0\\
0&0&0&0&0&0&0&1\\
1&0&0&0&0&0&0&0\\
0&-1&0&0&0&0&0&0\\
0&0&-1&0&0&0&0&0\\
0&0&0&-1&0&0&0&0
\end{array}\right),\\
\Gamma_5 = \left(\begin{array}{cccccccc}
0&0&0&0&0&0&0&-1\\
0&0&-1&0&0&0&0&0\\
0&1&0&0&0&0&0&0\\
0&0&0&0&-1&0&0&0\\
0&0&0&1&0&0&0&0\\
0&0&0&0&0&0&1&0\\
0&0&0&0&0&-1&0&0\\
1&0&0&0&0&0&0&0
\end{array}\right),
& \Gamma_6=\left(\begin{array}{cccccccc}
0&0&0&0&0&0&1&0\\
0&0&0&-1&0&0&0&0\\
0&0&0&0&1&0&0&0\\
0&1&0&0&0&0&0&0\\
0&0&-1&0&0&0&0&0\\
0&0&0&0&0&0&0&1\\
-1&0&0&0&0&0&0&0\\
0&0&0&0&0&-1&0&0
\end{array}\right),\\
\Gamma_7 = \left(\begin{array}{cccccccc}
0&1&0&0&0&0&0&0\\
-1&0&0&0&0&0&0&0\\
0&0&0&0&0&0&0&1\\
0&0&0&0&0&0&-1&0\\
0&0&0&0&0&1&0&0\\
0&0&0&0&-1&0&0&0\\
0&0&0&1&0&0&0&0\\
0&0&-1&0&0&0&0&0
\end{array}\right).
\end{array}\nn
\end{align}}

\end{document}